\documentclass[%
 aip,
 sd,%
 jcp,
 amsmath,amssymb,
 reprint,%
]{revtex4-1}

\usepackage{graphicx}% Include figure files
\usepackage{dcolumn}% Align table columns on decimal point
\usepackage{bm}% bold math
\begin{document}

\preprint{AIP/123-QED}

\title{Effective diffusion coefficient including the Marangoni effect}% Force line breaks with \\
%\thanks{Footnote to title of article.}

\author{Hiroyuki Kitahata}
\email{kitahata@chiba-u.jp.}
\affiliation{Department of Physics, Chiba University, Chiba 263-8522, Japan}

\author{Natsuhiko Yoshinaga}%
\affiliation{WPI-AIMR, Tohoku University, Sendai, Miyagi 980-8577, Japan}%
\affiliation{MathAM-OIL, AIST, Sendai, Miyagi 980-8577, Japan}

\date{\today}% It is always \today, today,

\begin{abstract}
Surface-active molecules supplied from a particle fixed at the water surface create a spatial gradient of the molecule concentration, resulting in Marangoni convection. Convective flow transports the molecules far from the particle, enhancing diffusion. We analytically derive the effective diffusion coefficient associated with the Marangoni convection rolls. The resulting estimated effective diffusion coefficient is consistent with our numerical results and the apparent diffusion coefficient measured in experiments. 
\end{abstract}

%05.45.-a Nonlinear dynamics and chaos  
%47.55.dk Surfactant effects 
%82.40.Ck Pattern formation in reactions with diffusion, flow and heat transfer  
%\keywords{Marangoni effect, surface tension, camphor, effective diffusion}%Use showkeys class option if keyword

\maketitle

\section{Introduction}

Self-propelled active materials have attracted increasing attention as a method for understanding biological systems from the viewpoint of physics.\cite{Mikhailov,Ramaswarmy,Vicsek} The studies in this field include not only real biological systems {\it in vivo} and {\it in vitro} but also synthesized physico-chemical systems. Physico-chemical systems are advantageous because the parameters can be controlled and specifically designed in order to clarify the desired mechanism in a quantitative manner. The camphor-water system is one of the most well-studied physico-chemical systems for self-propulsion.\cite{camphor,Nakata,Hayashima,PhysicaD,Grzybowski,Grzybowski2,Nishimori1,Nishimori2,PCCPreview,Koyano1,Koyano2,Nagayama,Suematsu} Camphor is a volatile organic material, which reduces the surface tension of water. When a camphor particle is placed onto the water surface, camphor molecules spread at the water surface, inducing a decrease in the surface tension around the camphor particle. The surface tension gradient at the water surface can drive the motion of the camphor particle itself.

A mathematical model for camphor motion, which is composed of a partial differential equation for the surface concentration of camphor particles, has been proposed by Nagayama et al.\cite{Nagayama} The model is simple, mathematically tractable, and reproduces the self-propulsion of camphor particles. The model predicts the transition between a stationary and
self-propelled states, and also the speed of the self-propulsion as a function of physico-chemical parameters, such as a diffusion coefficient of chemical molecules. Recently, considerable experimental efforts have been made to determine the parameters in the model,\cite{Nagayama} such as the sublimation rate, friction constant, and supply rate of camphor molecules from a camphor particle. Most of the results support the prediction of the theoretical model, but it was found that the estimated diffusion coefficient for the camphor molecules at the surface is approximately $10^{-3}~{\rm m}^2\cdot{\rm s}^{-1}$.\cite{Suematsu} Under equilibrium conditions, however, the diffusion coefficient of the molecules should be on the order of $10^{-9}~{\rm m}^2\cdot{\rm s}^{-1}$, which is six orders of magnitude smaller than the observed value. The purpose of this work is to understand this discrepancy.
We suggest that the Marangoni effect, which drives flow at the surface under the existence of the surface tension gradient,\cite{Marangoni,Marangoni2} plays an important role. In other words, the apparent diffusion coefficient observed in experiments corresponds to the effective diffusion coefficient enhanced by the Marangoni effect. In fact, Marangoni convection was experimentally observed in the camphor-water system.\cite{PCCP,SDS}

In the present article, we theoretically investigate the dynamics of the concentration profiles of camphor molecules and the flow profile of the aqueous phase.
Here, we assume the situation that a camphor particle is fixed at a certain position. The camphor molecules are dissolved from the camphor particle, and Marangoni flow occurs around the particle.
We show that the effect of the flow in the aqueous phase may be described as the effective diffusion coefficient, which depends on the wave number.\cite{Forster} The enhancement of the diffusion coefficient is understood as the effect of convective flow associated with the scale of the Marangoni convection. Numerical calculations based on the Navier-Stokes equation are compared with the analytical results. As a result, we obtain that the effect of the Marangoni flow cannot be rigorously represented as an effective diffusion coefficient, but approximately can be represented by considering the effective diffusion coefficient depending on the wave number.

Although mass transport by flow has a long history,\cite{Leal} surfactant transport by the Marangoni flow self-generated by a surface tension gradient has less been studied. We are aware of theoretical calculation by Bratukhin and Maurin,\cite{Bratukhin:1968} which was not often cited. The interesting issue in this phenomenon is that the flow is generated by inhomogeneity of a surfactant concentration through the surface tension, and in turn, the flow modifies the distribution of surfactants. Recently, several experiments have been performed to clarify the generic aspect of this system.\cite{Roche:2014,LeRoux:2016,Mandre:2017,Mandre:2017b} Specifically, it was found that for the large P{\'{e}}clet number and the finite Reynolds number, the velocity field maintains self-similar profiles. We consider in this this study a different situation where the velocity generated by the Marangoni flow is sufficiently slow, i.e., the Reynolds number is small (see Sec.\ref{sec.nond.num}).

\section{Model}

\begin{figure}
\begin{center}
\includegraphics{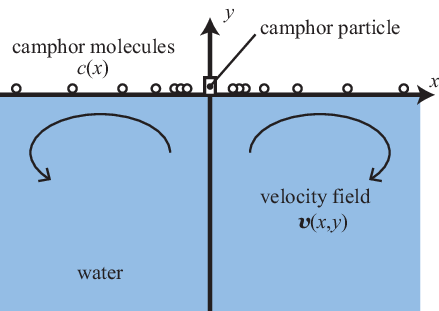}
\end{center}
\caption{Schematic illustration of the considered system. The $x$-axis corresponds to the water surface, and the $y$-axis is set in the vertical direction. We consider a camphor particle to be a point particle fixed at the origin.}
\label{fig1}
\end{figure}

We consider a two-dimensional camphor-water system, where the $x$-axis corresponds to the water surface, and the $y$-axis is in the vertical direction (Fig.~\ref{fig1}). We set $c(x)$ as the surface concentration of camphor molecules and $\bm{v}(x,y) = v_x(x,y) \bm{e}_x + v_y(x,y) \bm{e}_y$ as the flow profile in the aqueous phase, where $\bm{e}_x$ and $\bm{e}_y$ are the unit vectors in the $x$ and $y$ directions, respectively. The $x$-component of the flow velocity at the surface is set as $V(x)$, that is, $V(x) = v_x(x,0)$. The time evolution equation for $c$ is written as
\begin{equation}
\frac{\partial c}{\partial t} + \frac{\partial}{\partial x} \left(V c\right) = D \frac{\partial^2 c}{\partial x^2} - a c + f_0 \delta(x), \label{c-evo}
\end{equation}
where $D$ is the diffusion coefficient of camphor molecules under equilibrium condition, $a$ is the sublimation rate, and $f_0$ is the supply rate of camphor molecules from the particle. The flow field $\bm{v}$ obeys the Navier-Stokes equation
\begin{equation}
\rho \left( \frac{\partial}{\partial t} + \bm{v} \cdot \nabla \right) \bm{v} = - \nabla p + \eta \nabla^2 \bm{v}, \label{NS}
\end{equation}
where $p$ is the pressure, and $\rho$ and $\eta$ are the density and viscosity of the fluid, respectively. In addition, we assume incompressibility: 
\begin{equation}
\nabla \cdot \bm{v} = 0. \label{incompressible}
\end{equation}
The surface tension is a decreasing function of the camphor surface concentration. For simplicity, we assume a linear relation between the surface tension $\gamma$ and the surface concentration of camphor $c$ as
\begin{equation}
\gamma = \gamma_0 - \Gamma c, \label{surfacetension}
\end{equation}
where $\Gamma~(>0)$ is a proportionality constant, and $\gamma_0$ is the surface tension of pure water. The surface tension gradient induces flow through the boundary condition as
\begin{equation}
\eta \left. \frac{\partial v_x}{\partial y}\right|_{y=0} = \frac{\partial \gamma}{\partial x}, \label{boundarycondition}
\end{equation}
which is derived from the stress balance at the surface.\cite{interface,Landau,Young} In this theoretical analysis, we apply the Stokes approximation; that is, we neglect the inertia term $\rho (\bm{v} \cdot \nabla) \bm{v}$ in the Navier-Stokes equation.\cite{Brenner}
The justification of this approximation will be discussed later. 

Here, we consider the steady state; $\partial c/ \partial t=0$, and $\partial \bm{v}/ \partial t= \bm{0}$. The flow profile can be written as a functional of the surface tension profile $\gamma (x)$. When $\gamma(x)$ can be expanded in Fourier space,
\begin{equation}
\gamma(x) = \gamma_{{\rm c}0} + \int_0^\infty \left( \gamma_{\rm c}(k) \cos k x + \gamma_{\rm s}(k) \sin kx \right) {\rm d}k,
\end{equation}
the flow velocity at the surface $V(x)$ is described as
\begin{equation}
V(x) = \frac{1}{2\eta} \int_0^\infty \left( - \gamma_{\rm c}(k) \sin kx + \gamma_{\rm s}(k) \cos kx \right) {\rm d}k. \label{convection}
\end{equation}
The derivation of Eq.~(\ref{convection}) is shown in Appendix \ref{app-a}. 
 
Without the effect of the Marangoni flow, the concentration field is obtained as the solution of
\begin{equation}
D \frac{\partial^2 c}{\partial x^2} - ac + f_0 \delta(x) = 0. \label{stable}
\end{equation}
That is,
\begin{equation}
c(x) = \frac{f_0}{2 \sqrt{aD}} \exp \left(- \sqrt{\frac{a}{D}} \left|x \right| \right).
\end{equation}
It is noted that this steady-state concentration profile can be described in the form of Fourier transformation:
\begin{equation}
c(x) = \frac{f_0}{\pi a} \int_0^\infty \frac{1}{1 + D k^2 / a}\cos kx {\rm d}k.
\end{equation}

\begin{figure}
\begin{center}
\includegraphics{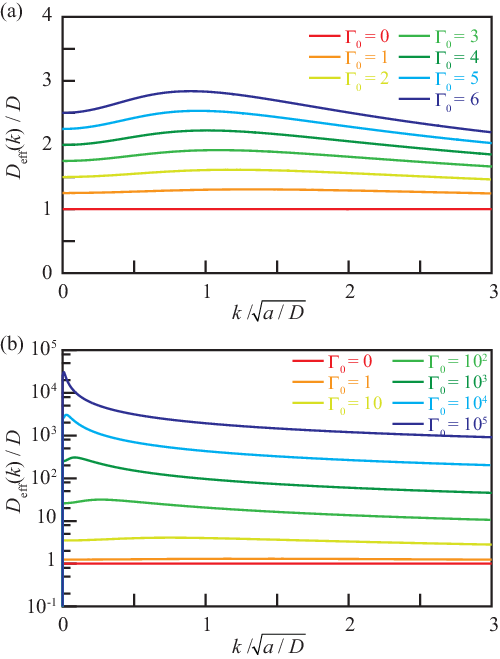}
\end{center}
\caption{Plot of $D_{\rm eff}(k)/D$ against $k$ for various $\Gamma_0$. (a) Linear plot for small $\Gamma_0$. (b) Semi-logarithmic plot for large $\Gamma_0$.}
\label{fig2}
\end{figure} 
With the effect of the flow, we cannot solve Eq.~(\ref{c-evo}) analytically because the velocity field at the surface $V(x)$ is dependent on the concentration field, and thus the advection term is nonlinear. Therefore, approximation is necessary.
When the camphor particle stops at the origin, both the concentration and flow profile should be symmetric, reflecting the symmetry of the system. If this symmetric decaying profile is expressed as an exponential function, the effective diffusion coefficient is determined by the characteristic length of the exponential function. We express the concentration profile $c(x)$ expanded in the Fourier space as
\begin{equation}
c(x) = \frac{f_0}{\pi a} \int_0^\infty \frac{1}{1 + D_{\rm eff}(k) k^2 / a} \cos kx {\rm d}k. \label{c-profile}
\end{equation}
Once we know the concentration field including the effect of Marangoni flow, for example, in experiments or in numerical simulations, Eq.~(\ref{c-profile}) gives exact form of $k$-dependent diffusion coefficient. As mentioned above, the explicit form of the solution of Eq.~(\ref{c-evo}) is not available, and $D_{\rm eff}(k)$ should be approximated. Our strategy is to evaluate the nonlinear advection term in Eq.(\ref{c-profile}) under the assumption that $D_{\rm eff}(k)$ is constant in $k$. Then, we plug the evaluated advection term into Eq.~(\ref{c-evo}), solve the equation for $D_{\rm eff}(k)$, and find a representative wave number, which is consistent condition with the above assumption. The consistent condition is extrema of $D_{\rm eff}(k)$ as a function of $k$. This method enables us to estimate an effective diffusion coefficient as well as the size of convective rolls, which does not appear in the standard perturbative expansion (see Sec.~\ref{sec.comparison}). The disadvantage of the current method is that it is under uncontrolled approximation. Therefore, in Sec.~\ref{sec.numerical}, we check the validity of the method by comparing $D_{\rm eff}(k)$ in our theoretical calculation and numerical results.

\section{Effective Diffusion Coefficient}

In the presence of advection, Eq.~(\ref{c-evo}) cannot be solved analytically. Nevertheless, we may use an ansatz where the concentration profile $c(x)$ is expanded as in Eq.~(\ref{c-profile}), using the effective diffusion coefficient $D_{\rm eff}(k)$ that depends on wave number $k$.\cite{Forster} We have used the fact that both the concentration and flow profiles are symmetric when the camphor particle stops at the origin. By substituting Eq.~(\ref{c-profile}) into Eq.~(\ref{c-evo}) with $\partial c/\partial t = 0$ and using Eq.~(\ref{convection}), we may obtain $D_{\rm eff}(k)$ self-consistently, under the assumption that $D_{\rm eff}(k)$ does not depend on $k$ in the calculation of the integration. By defining the following nondimensionalized parameter
\begin{equation}
\Gamma_0 = \frac{\Gamma f_0}{\pi \eta D a}, \label{gamma0}
\end{equation}
the equation for determining the effective diffusion coefficient becomes (see Appendix~\ref{appendixB})
\begin{equation}
\frac{ \Gamma_0}{2} G\left(\sqrt{\frac{D_{\rm eff}(k) k^2}{a}}\right)  = \frac{D_{\rm eff}(k)}{D} - 1, \label{eq17}
\end{equation}
where
\begin{equation}
 G(\xi) = \frac{(1 + \xi^2) \left(\xi \arctan \xi + \ln (1 + \xi^2)\right)}{ \xi^2 \left(4 + \xi^2 \right)}. \label{Gxi}
\end{equation}
The plot of the nondimensionalized effective diffusion coefficient $D_{\rm eff}(k)/D$ as a function of the nondimensionalized wave number $k/\sqrt{a/D}$ is shown in Fig.~\ref{fig2}. 

For small wave numbers, $\lim_{\xi \rightarrow 0} G(\xi) = 1/2$, and the effective diffusion coefficient becomes
\begin{equation}
D_{\rm eff}(0) = D \left( 1 + \frac{\Gamma_0}{4} \right).
\end{equation}
Another property of the effective diffusion coefficient is a shift of the peak as shown in Fig.~\ref{fig2}. In fact,  the wave number $k_{\rm max}$ at which $D_{\rm eff}$ has a maximum value goes to $+0$ when $\Gamma_0$ goes to infinity. This behavior occurs because $k_{\rm max}^2 D_{\rm eff}(k_{\rm max}) / a = \xi_{\rm max} = {\rm const.}$, where $\xi_{\rm max}$ is the positive root of ${\rm d}G/{\rm d}\xi = 0$. Here, $\xi_{\rm max}$ is numerically estimated as $\simeq 1.50856$. The effective diffusion coefficient $D_{\rm eff}(k)$ at $k=k_{\rm max}$ can be calculated as
\begin{equation}
 D_{\rm eff}(k_{\rm max}) = D (1 + \alpha \Gamma_0),
  \label{Deff.max}
\end{equation}
where the constant $\alpha = G(\xi_{\rm max})/2$ is evaluated as $\alpha \simeq 0.306554$.

The ratio between the diffusion coefficient and sublimation rate sets a
length scale $\lambda=\sqrt{D/a}$, whose value will be discussed in Sec~\ref{sec.comparison}. For the small wave length limit, i.e., $k \lambda \rightarrow \infty$, $D_{\rm eff} / D  -1 \rightarrow 0$ because $G(\xi) \sim 1/\xi$ as $\xi\rightarrow +\infty$. Expanding $D_{\rm eff}$ around this limit as $D_{\rm eff} = D(1 + f(k))$, we obtain
\begin{equation}
\frac{\Gamma_0}{2} \sqrt{\frac{a}{D}} \frac{1}{k \sqrt{1 + f(k)}} = f(k).
\end{equation}
Because $f(k)$ is infinitesimally small when $k$ is sufficiently large, we obtain
\begin{equation}
D_{\rm eff}(k) \sim D + \frac{\Gamma_0}{2} \sqrt{\frac{a}{D}} \frac{1}{k}.
\end{equation}
This means that the gradient of the concentration field around the camphor particle is not significantly affected by the convective flow.

We now consider the diffusion-like phenomenon at the large spatial scale. For this purpose, $D_{\rm eff}(k)$ for large $k$ is not suitable. Instead, it is natural to take the effective diffusion coefficient as $D_{\rm eff} \sim D_{\rm eff}(k_{\rm max}) = D (1 + \alpha \Gamma_0)$. This is because the diffusion is dominated by the convective roll structure, whose length scale is associated with$k_{\rm max}$ as $2\pi/ k_{\rm max}$.

\section{Numerical calculation}
\label{sec.numerical}

\begin{figure}
\begin{center}
\includegraphics{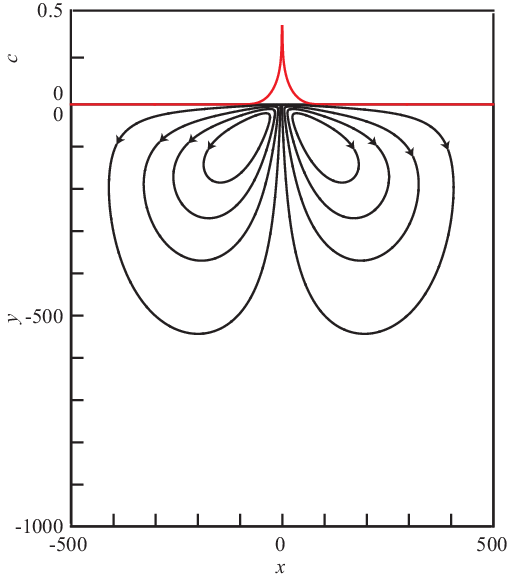}
\end{center}
\caption{Stationary concentration and flow profiles represented by streamlines for $\Gamma_0 = 100$, obtained by numerical calculation.}
\label{fig3}
\end{figure} 

\begin{figure}
\begin{center}
\includegraphics{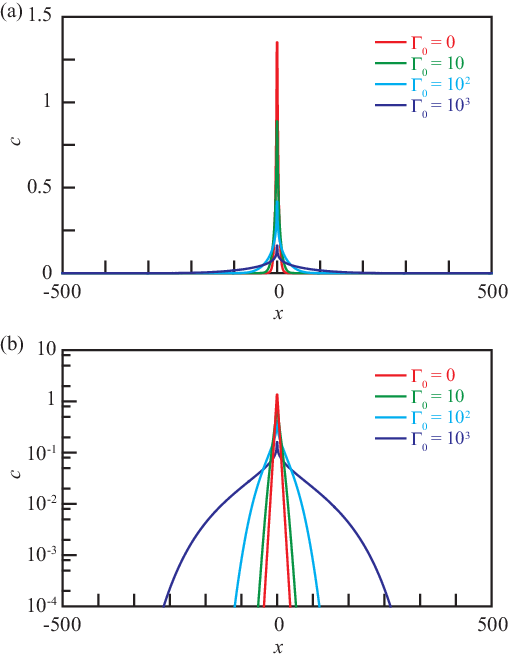}
\end{center}
\caption{Stationary concentration profile of $c$ obtained from
 numerical calculations for each $\Gamma_0$: (a) linear plot, and (b) semi-logarithmic plot.}
\label{fig4}
\end{figure} 

To confirm the validity of the present approach, we performed numerical calculations. The reaction-diffusion equation Eq.~(\ref{c-evo}) is calculated using the Euler method with Neumann boundary conditions at $x = \pm L_x$. The hydrodynamics, given in Eqs.~(\ref{NS}) and (\ref{incompressible}) are solved by the stream function-vorticity method supplemented with the boundary conditions in Eq.~(\ref{boundarycondition}) at $y=0$ and non-slip boundary conditions for $\bm{v}$ at $x = \pm L_x$ and $y = -L_y$. In the theoretical analysis, we consider an infinite half plane, but we adopt a sufficiently large plane, $-L_x \leq x \leq L_x$ and $-L_y \leq y \leq 0$, in the numerical calculation. We set $L_x = 500$ and $L_y = 1000$ so that the system size does not affect the results. We set the time step as $\Delta t = 10^{-4}$, and the space grid as $\Delta x = 1$. The numerical calculations were performed until the concentration and flow profiles approximately reach stationary solutions. The representative steady-state concentration and flow profiles are shown in Fig.~\ref{fig3}. We calculated results for various nondimensionalized parameters of $\Gamma_0$ by changing the parameter $\Gamma$. The other parameters were set as $D = 1$, $a = 0.1$, $f_0 = 1$, $\rho = 1$, and $\eta = 1000$. The delta function for the supply of the camphor molecules in Eq.~(\ref{c-evo}) is approximated by the supply at two discretized points. The steady-state concentration fields of $c$ for various $\Gamma_0$ are shown in Fig.~\ref{fig4}(a) on a linear scale and in Fig.~\ref{fig4}(b) on a logarithmic scale.

\begin{figure}
\begin{center}
\includegraphics{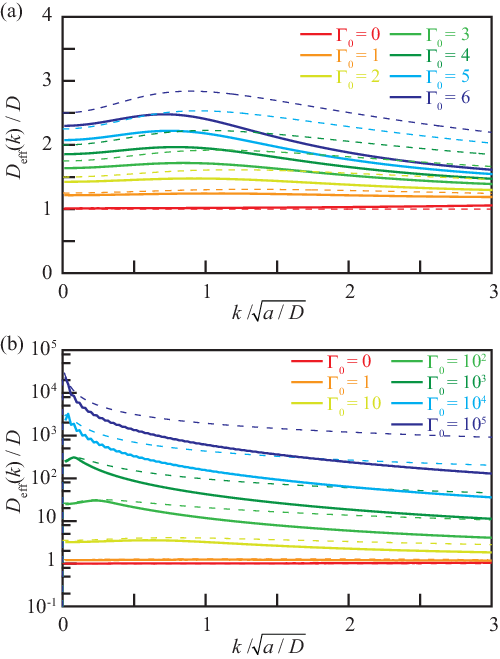}
\end{center}
\caption{Numerical results of $D_{\rm eff}(k) / D$ as a function of $k / \sqrt{a/D}$ for each $\Gamma_0$. Stationary concentration profile of $c$ for each $\Gamma_0$ was obtained, from which $D_{\rm eff}/D$ was calculated: (a) linear plot for small $\Gamma_0$ and (b) semi-logarithmic plot for large $\Gamma_0$. The corresponding analytical results in Fig.~\ref{fig2} are shown by the dashed curves.}
\label{fig5}
\end{figure}

The normalized effective diffusion coefficient $D_{\rm eff}/D$ as a function of the normalized wave number $k / \sqrt{a/D}$ is obtained from the results of the numerical calculation using the Fourier cosine transformation. The results are shown in Fig.~\ref{fig5}, where the numerical results are shown by solid curves and the analytical results by dashed curves. 
Both results show qualitatively same tendency and, in particular, the
 analytical results reproduce a peak at the specific wave number.
 Nevertheless, we found there were quantitative discrepancies between the
 numerical and analytical results for large $\Gamma_0$. This is because the assumption that the
 $D_{\rm eff}$ is independent of $k$ is broken for large $\Gamma_0$.
 As a result, the evaluation of the nonlinear advection term in Eq.~(\ref{c-evo}) and the left-hand side of Eq.~(\ref{convection}) become inaccurate.

Other sources of the discrepancies are considered to be a result of the
following aspects: (i) discretization in the numerical calculation, (ii) system size of the numerical calculation, and (iii) higher-order wave number dependence on the effective diffusion coefficient. The effect of the discretization was important for the discrepancy at high-$k$ region. Even when $\Gamma_0 = 0$, there remains some discrepancy in Fig.~\ref{fig5} in high-$k$ region. This also implies that the discretization affected the discrepancy. On the while, the numerical calculation showed that the effect of the system size did not affect so significantly. As for the third aspect, our analysis is valid around ${\rm d} D_{\rm eff}/{\rm d}k \ll 1$. As $\Gamma_0$ is increased, such dependence becomes stronger and causes a worse estimate away from the peak value. Although there are some discrepancies between theoretical prediction and numerical results, the important point is that the profile of $D_{\rm eff}(k) / D$ has a peak close to $k = k_{\rm max}$, which was reproduced by numerical calculation. Therefore we suppose that our theoretical derivation reflects the essential features of the dynamic of molecule transport at water surface.
Details on the check with numerical calculation are shown in Appendix~\ref{app-validity}.

\section{Comparison with the experimental results\label{sec.comparison}}

Here, we discuss the validity of our estimation based on the
experimental results.\cite{Suematsu} The rate of sublimation to air and dissolution to the aqueous phase $a$, the supply rate of camphor molecules from the camphor grain $f_0$,
and the proportionality constant between the surface tension and the camphor concentration, $\Gamma$, are estimated to be $a \sim 10^{-2}~{\rm s}^{-1}$, $f_0 \sim 10^{-11}~{\rm mol}\cdot{\rm s}^{-1}$, and $\Gamma \sim 10^{-3}~{\rm Pa}\cdot{\rm s}$, respectively. The values of $a$ and $f_0$ were estimated by the measurement of the relaxation process of the surface pressure when a camphor boat is put onto or removed from the water surface, together with the measurement of the weight change of the camphor disk after it moves around at the water surface. In the experiments,\cite{Suematsu} a camphor boat was used in the place of a camphor disk but it is expected that the orders of magnitude of these values are the same. $\Gamma$ is theoretically derived as $\Gamma = R T$, where $R$ is the gas constant ($R \simeq 8.31~{\rm J} \cdot {\rm mol}^{-1} \cdot {\rm K}^{-1}$) and $T$ is absolute temperature ($T \simeq 300~{\rm K}$), considering Gibbs adsorption isotherm and Henry isotherm.\cite{Suematsu,Chang} The viscosity of water $\eta$, and the diffusion coefficient of the camphor molecule $D$, are known as $\eta \sim 10^{-3}~{\rm Pa}\cdot{\rm s}$ and $D\sim 10^{-9}~{\rm m}^2\cdot {\rm s}^{-1}$, respectively, at room temperature. 
The unit length is then estimated as $\lambda \sim 10^{-3}$~m.
From these values, the nondimensionalized number $\Gamma_0$ is estimated
as
\begin{equation}
\Gamma_0 \simeq \frac{2.5 \times 10^3 [{\rm J}\cdot {\rm mol}^{-1}] \cdot 10^{-11}[{\rm mol} \cdot {\rm s}^{-1}]}{3.14 \cdot 10^{-3} [{\rm Pa} \cdot {\rm s}] \cdot 10^{-9} [{\rm m}^2 \cdot {\rm s}^{-1}] \cdot 10^{-2}[{\rm s}]} \sim 10^6.
\end{equation}
With this value, the effective diffusion coefficient $D_{\rm eff}$ is calculated as
\begin{equation}
D_{\rm eff} \sim D ( 1 + \alpha \Gamma_0) \sim 10^{-3}~[{\rm m}^2 \cdot {\rm s}^{-1}].
\end{equation}
In the previous paper, the apparent diffusion coefficient was estimated as $4 \times 10^{-3}~{\rm m}^2\cdot{\rm s}^{-1}$,\cite{Suematsu} which is consistent with our theoretical estimation.

It should be noted that we consider the two-dimensional system in the theoretical analysis and numerical calculation, while the experimental system is a three dimensional one. The profile of the Marangoni flow structure might be different, but it might be worthwhile to compare the order of the effective diffusion coefficient. Extension of our framework to the three-dimensional system remains as future work.

\section{Nondimensional numbers}
\label{sec.nond.num}

In this section, we consider the relationship between $\Gamma_0$ and
other nondimensional numbers.\cite{Leal}
The Reynolds number ${\rm Re} = \rho R U / \eta$ is
\begin{equation}
{\rm Re} \sim \frac{\pi \rho \Gamma f_0}{\eta^2 a},
\end{equation} 
from the characteristic length $R$,
\begin{equation}
R \sim x_0 = \sqrt{\frac{D}{a}}.
\end{equation}
The characteristic velocity $U$ is
\begin{equation}
U \sim \frac{\Gamma f_0}{\eta a x_0} \sim \frac{\Gamma f_0}{\pi \eta \sqrt{D a} }. 
\end{equation}
The characteristic velocity is obtained from the mechanical balance at the surface as in Eq.~(\ref{boundarycondition}): 
\begin{equation}
\eta \frac{U}{R} \sim \frac{\gamma}{R} \sim \frac{\Gamma c}{R}.
\end{equation}
The Reynolds number can also be described using the nondimensional constant $\Gamma_0$ defined in Eq.~(\ref{gamma0}) as
\begin{equation}
 {\rm Re} \sim \frac{\rho \Gamma f_0}{\eta^2 a} = \frac{\pi \Gamma_0}{{\rm Sc}},
\end{equation}
where ${\rm Sc}$ is Schmidt number, i.e., ${\rm Sc} = \eta / (D \rho)$. The nondimensional number $\Gamma_0$ is nothing but the Peclet number, because $ {\rm Pe} = {\rm Re} {\rm Sc}$,
\begin{equation}
\Gamma_0 = \frac{\rm Pe}{\pi}.
\end{equation}
\begin{figure}
\begin{center}
\includegraphics[scale=0.95]{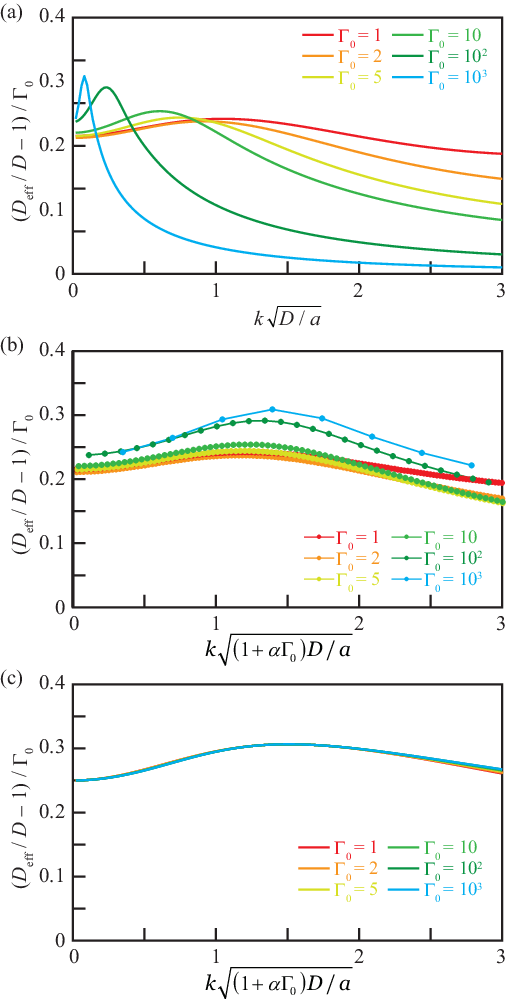}
\end{center}
\caption{(a) Numerical results of $(D_{\rm eff} / D - 1) / \Gamma_0$ against $k \sqrt{D/a}$ for various $\Gamma_0$. The peak positions strongly depend on $\Gamma_0$. (b) Numerical results of $(D_{\rm eff} / D - 1) / \Gamma_0$ against $k \sqrt{(1 + \alpha \Gamma_0)D/a}$ for various $\Gamma_0$. The peak positions are almost independent of $\Gamma_0$ except for the data with $\Gamma_0 = 10^4$. The number of data points for large $\Gamma_0$ is small due to the rescaling effect. (c) Analytical results of $(D_{\rm eff} / D - 1) / \Gamma_0$ against $k \sqrt{(1 + \alpha \Gamma_0)D/a}$ for various $\Gamma_0$. The curves almost collapse to a universal curve.}
\label{fig6}
\end{figure}

The Schmidt number of water is calculated to be
\begin{equation}
{\rm Sc} \simeq \frac{ 10^{-3}~[{\rm Pa \cdot s}]}{10^{-9}~[{\rm m^2/s}] \cdot 10^3~{[{\rm kg / m^3}]}}\sim 10^3,
\end{equation}
where we use $D \sim 10^{-9}~{\rm m^2/s}$, $\rho \sim 10^3~{\rm kg / m^3}$, and $\eta = 10^{-3}~{\rm Pa \cdot s}$. The Schmidt number in the numerical calculation was also set to be 1000 by setting $\rho = D = 1$ and $\eta = 1000$. A large Schmidt number means the nonlinear term, $\rho (\bm{v} \cdot \nabla) \bm{v}$, in the Navier-Stokes equation in Eq.~(\ref{NS}), is insignificant compared with the nonlinear term, $\bm{v} \cdot \nabla c$, in the evolution equation for the concentration in Eq.~(\ref{c-evo}). In numerical calculation, we have checked the nonlinear term in the Navier-Stokes equation does not matter much as shown in Appendix~\ref{app-validity}. In the recent experimental observation,\cite{Kitahata:2017} the camphor disk is placed away from the water surface in order to realize a smaller supply of surfactants.
In this case, the Reynolds number is as small as ${\rm Re} \simeq 1$.
Our theoretical calculation is suited to this system.

Note that our model is different from the previous works
\cite{Bratukhin:1968,Mandre:2017,Mandre:2017b} by the evaporation
effect.
This is manifested by the linear damping term of $c$ in Eq.~(\ref{c-evo}).
Because of this term, the self-similar profile is prohibited.
This evaporation effect might weaken the effect of inertia term in
the Navier-Stokes equation.

\section{Comparison with Perturbative Expansion}
\label{sec.experiment}

To discuss the meaning of our derivation of $D_{\rm eff}$, we compare it with the perturbation method. The diffusion coefficient under the perturbation method $D'$ is
\begin{equation}
 \frac{D'}{D} = 1 + \frac{\Gamma_0}{2} G(x_0 k) + \mathcal{O}(\epsilon^2). \label{perturbation}
\end{equation}
The detailed derivation is shown in Appendix~\ref{app-perturbation}. This result looks similar to Eq.~(\ref{Deff.max}) in that both have additional terms proportional to $\Gamma_0$ whose coefficient is proportional to the wave number $k$. However, in Eq.~(\ref{Deff.max}), the wave number $k_{\rm max}$ for the maximum $D_{\rm eff}$ depends on $\Gamma_0$ while $D'(k)$ in Eq.~(\ref{perturbation}) has a maximum value at a constant $k = \xi_{\rm max}/x_0$ even when $\Gamma_0$ changes. 

Comparing Eqs.~(\ref{eq17}) and (\ref{perturbation}), our method allows $x_0$ in Eq.~(\ref{perturbation}) to be dependent on the wave number $k$. The physical meaning of this is the rescaling of the wave number. To obtain an effective diffusion coefficient for enhanced diffusion due to Marangoni convection, it is natural to take the value with $\partial D_{\rm eff}/ \partial k = 0$, which is equivalent to the peak value of $D_{\rm eff}$. Therefore, we used the effective diffusion coefficient given in Eq.~(\ref{Deff.max}).

To confirm the validity of the present method, we plotted the numerical results with rescaled axes. The wave number $k$ should be rescaled as $k \sqrt{D_{\rm eff} / a}$. Considering that $D_{\rm eff}$ is approximately described by $D_{\rm eff} = D(1 + \alpha \Gamma_0)$, we rescale $k$ as $k \sqrt{1 + \alpha \Gamma_0}$. In Fig.~\ref{fig6}, $(D_{\rm eff} / D - 1) / \Gamma_0$ obtained from the numerical calculation is plotted against $k \sqrt{D/a}$ in Fig.~\ref{fig6}(a) and against $k \sqrt{(1 + \alpha \Gamma_0)D/a}$ Fig.~\ref{fig6}(b). 
The plots in Fig.~\ref{fig6}(a) do not collapse because the peak
positions strongly depend on $\Gamma_0$.
On the other hand, the peak position in Fig.~\ref{fig6}(b) with the
rescaled wave number tends to collapse to a universal curve.
This suggests the validity of our analysis of the characteristic wave number.
In Fig.~\ref{fig6}(c), $(D_{\rm eff} / D - 1) / \Gamma_0$ obtained from the analytical calculation is plotted against $k\sqrt{(1 + \alpha \Gamma_0)D/a}$. The curves approximately collapse to a universal curve. These results show that our method is applicable for diffusion phenomena at the scale of the size of the convective roll.
Comparing the theoretical and numerical results shown in Figs.~\ref{fig6}(b) and (c), there are some discrepancies on the values of $(D_{\rm eff}/D-1)/\Gamma_0$ near $k \sim 0$. In Fig.~\ref{fig6}(b), we can see some discrepancies between the curves with different $\Gamma_0$. We consider these are due to the discretization effect. For accurate numerical calculation for wide range of $k$, we have to use small spatial step. Since the concentration profile is localized, the discretization step is insufficient and we cannot obtain the accurate value. In fact, we changed the grid size in numerical calculation and and confirmed that the numerical results approach the theoretical results in Appendix~\ref{app-validity}.

\section{Summary}

When a surface-active chemical compound is supplied from a particle placed at the water surface, Marangoni convection is induced, which accelerates the transport of the surface active chemical compounds. This process can be described using the effective diffusion coefficient, which is derived analytically in this paper. We derived the effective diffusion coefficient under the approximation of weak Marangoni convection, the estimated value is compatible with previously reported experimental results.\cite{Suematsu} The mathematical approach for strong Marangoni convection remains as future study.

In the present calculation, we consider only a stationary camphor particle fixed at a certain position. Therefore, the effective diffusion coefficient obtained in this paper may be different from that for a spontaneously moving camphor particle. Nevertheless, we expect that the effective diffusion coefficient is valid when the system is near the bifurcation point from the rest state to motion, i.e., the particle is moving at a low velocity.\cite{Nagayama,Koyano1,Koyano2} In this case, the velocity of self-propulsion is perturbatively expanded and is expressed by the deviation of the concentration field under motion from the concentration field at the stationary state. Then, we may simply replace the bare diffusion coefficient by the effective one to compute the isotropic concentration field at the stationary state. The situation in which the particle is moving at a finite constant velocity is left for future work.

\begin{acknowledgments}

The authors thank Nobuhiko J. Suematsu (Meiji University, Japan), Alexander S. Mikhailov (Fritz-Haber Institute), and Yutaka Sumino (Tokyo University of Science, Japan) for their helpful discussion. This work was supported by JSPS KAKENHI Grants No.~JP15K05199, No.~JP26800219, No.~JP25103008, No.~JP26103503, No.~JP16H00793, and No.~JP17K05605.

\end{acknowledgments}

\appendix

\section{Derivaton of Eq.~(\ref{convection})}
\label{app-a}

We show the derivation of the steady state solution of Eqs.~(\ref{NS}) and (\ref{incompressible}) with the boundary condition in Eq.~(\ref{boundarycondition}) and $\bm{v} = \bm{0}$ at $y = -H$. The surface tension $\gamma$ is related to the concentration $c$ by Eq.~(\ref{surfacetension}). By defining the stream function $\psi$ as
\begin{align}
v_x =& \frac{\partial \psi}{\partial y}, \label{vx}\\
v_y =& - \frac{\partial \psi}{\partial x}, \label{vy}
\end{align}
the incompressibility given in Eq.~(\ref{incompressible}) is always satisfied. The Stokes equation, which is Eq.~(\ref{NS}) without the inertia term,
is rewritten as the following equation of $\psi$,
\begin{equation}
\nabla^2 \left(\nabla^2 \psi \right) = 0. \label{eq-2}
\end{equation}

We assume that the surface tension profile can be expanded in Fourier space as
\begin{equation}
\gamma(x) = \gamma_{{\rm c}0} + \int_0^\infty \left(\gamma_{\rm c}(k) \cos kx + \gamma_{\rm s}(k) \sin kx \right) {\rm d}k.
\end{equation}
Because the equation is linear, the general solution satisfying periodicity in the $x$-direction is obtained as
\begin{widetext}
\begin{align}
\psi =& \left[ A_{\rm c}(k) e^{ky} + B_{\rm c}(k) e^{-ky} + C_{\rm c}(k) ye^{ky}+ D_{\rm c}(k) ye^{-ky} \right] \cos kx \nonumber \\
&+ \left[ A_{\rm s}(k)  e^{ky} + B_{\rm s}(k) e^{-ky} + C_{\rm s}(k) ye^{ky} + D_{\rm s}(k) ye^{-ky} \right]\sin kx, \label{gen}
\end{align}
\end{widetext}
where $A_{\rm c}(k)$, $B_{\rm c}(k)$, $C_{\rm c}(k)$, $D_{\rm c}(k)$, $A_{\rm s}(k)$, $B_{\rm s}(k)$, $C_{\rm s}(k)$, and $D_{\rm s}(k)$ are integration constants. Using the boundary conditions, we obtain
\begin{widetext}
\begin{align}
\psi = \frac{2kH^2 \sinh ky + 2kH y \sinh ky + y \cosh ky - y\cosh (2kH + ky)}{2\eta (2 k H - \sinh 2kH)}\left( -\gamma_{\rm c}(k) \sin kx + \gamma_{\rm s}(k) \cos kx \right),
\end{align}
\begin{align}
p = p_0 + \frac{2k^2 H \sinh ky - k \cosh (2kH + ky) + k \cosh ky}{2 k H - \sinh 2kH}\left(\gamma_{\rm c}(k) \cos kx + \gamma_{\rm s}(k) \sin kx \right),
\end{align}
\end{widetext}
where $p_0$ is a constant.

In this work, we consider the aqueous phase with an infinite depth, and take the limit of $H \rightarrow \infty$. Then, the stream function $\psi$ and the pressure $p$ converge to
\begin{align}
\psi = \frac{1}{2\eta} y e^{ky} \left( -\gamma_{\rm c}(k) \sin kx + \gamma_{\rm s}(k) \cos kx \right),
\end{align}
\begin{align}
p = p_0 + k e^{ky} \left( \gamma_{\rm c}(k) \cos kx + \gamma_{\rm s}(k) \sin kx \right).
\end{align}

By adding all modes, we obtain the stream function $\psi$, the pressure $p$, and the velocity field $(v_x, v_y)$ as follows: 
\begin{equation}
\psi = \frac{1}{2\eta} \int_0^\infty ye^{ky} \left(- \gamma_{\rm c}(k) \sin kx + \gamma_{\rm s}(k)\cos kx \right) {\rm d}k,
\end{equation}
\begin{equation}
p = p_0 + \int_0^\infty k e^{ky} \left( \gamma_{\rm c}(k)  \cos kx + \gamma_{\rm s}(k) \sin kx \right) {\rm d}k,
\end{equation}
\begin{equation}
v_x = \frac{1}{2\eta} \int_0^\infty (1 + ky)e^{ky} \left(- \gamma_{\rm c}(k) \sin kx + \gamma_{\rm s}(k) \cos kx  \right) {\rm d}k,
\end{equation}
\begin{equation}
v_y = \frac{1}{2\eta} \int_0^\infty k ye^{ky} \left( \gamma_{\rm c}(k) \cos kx + \gamma_{\rm s}(k) \sin kx \right) {\rm d}k.
\end{equation}

Finally, we determine the following flow velocity in the $x$-direction at the water surface: 
\begin{align}
V(x) &= v_x(x,0) \nonumber \\
& = \frac{1}{2\eta} \int_0^\infty \left(- \gamma_{\rm c}(k) \sin kx + \gamma_{\rm s}(k) \cos kx \right) {\rm d}k.
\end{align}

\section{Derivation of Eq.~(\ref{eq17})}
\label{appendixB}

The nonlinear term $\partial (V c) / \partial x$ is calculated by considering the coupling between two modes in Fourier space under the assumption that $D_{\rm eff}(k) = D_{\rm eff}$, i.e., independent of $k$.
\begin{align}
\frac{\partial}{\partial x}\left(V c\right) = & \frac{\Gamma {f_0}^2}{2\pi^2 \eta a^2} \frac{\partial}{\partial x} \left[\left( \int_0^\infty \frac{\sin kx}{1 + D_{\rm eff} k^2 / a}  {\rm d}k \right) \right. \nonumber \\ & \left. \times \left( \int_0^\infty \frac{\cos kx}{1 + D_{\rm eff} k^2 / a} {\rm d}k \right) \right] \nonumber \\
= \frac{\Gamma {f_0}^2}{2 \pi^2 \eta a^2} &\int_0^\infty F\left(\sqrt{\frac{D_{\rm eff} k^2}{a}}\right) k^2 \cos kx {\rm d}k, \label{modecoupling}
\end{align}
where
\begin{equation}
F(\xi) = \frac{\xi \arctan \xi + \ln (1 + \xi^2)}{ \xi^2 \left(4 + \xi^2 \right)}.
\end{equation}
Here, we use the following equality: 
\begin{align}
&\left(\int_0^\infty \frac{\sin kx}{1+{x_0}^2 k^2} {\rm d}k \right)\left(\int_0^\infty \frac{\cos kx}{1+{x_0}^2 k^2} {\rm d}k \right)\nonumber \\
&= \frac{1}{2} \int_0^\infty \left[ \int_0^k \frac{{\rm d}k'}{(1 + {x_0}^2{k'}^2)(1 + {x_0}^2(k - k')^2)}\right] \sin kx {\rm d}k \nonumber \\
&= \int_0^\infty \frac{x_0 k \arctan x_0k + \ln \left(1 + {x_0}^2k^2\right)}{{x_0}^2 k (4 + {x_0}^2k^2)} \sin kx {\rm d}k. 
\end{align} 
Therefore, by comparing the coefficient of $\cos kx$ in Eq.~(\ref{c-evo}), we obtain
\begin{equation}
\frac{ \Gamma_0}{2} F\left(\sqrt{\frac{D_{\rm eff}(k) k^2}{a}}\right) D
 k^2/a = - \frac{D k^2/a + 1}{D_{\rm eff}(k)k^2 / a+1} + 1. \label{B3}
\end{equation}
By multiplying both sides of Eq.~(\ref{B3}) with $(1 + D_{\rm eff}(k)k^2/a) / (Dk^2/a)$, and defining $G(\xi)$ as
\begin{equation}
G(\xi) = (1+ \xi^2) F(\xi),
\end{equation}
we lead Eqs.~(\ref{eq17}) and (\ref{Gxi}).

\section{Diffusion coefficient obtained with the perturbation method}
\label{app-perturbation}

We calculate the diffusion coefficient using the perturbation method. To do this, $\Gamma$ in Eq.~(\ref{surfacetension}) is treated as an infinitesimally small parameter $\epsilon$. The concentration field $c(x)$ is expanded with regard to $\epsilon$ as
\begin{equation}
c(x) = c_0(x) + \epsilon c_1(x) + \mathcal{O}({\epsilon^2}).
\end{equation}
In Fourier space, $\tilde{c}(k)$, the Fourier transform of $c(x)$, is also expanded with regard to $\epsilon$ as  
\begin{equation}
c(x) = \int_0^\infty \left( \tilde{c}_0(k) + \epsilon \tilde{c}_1(k) \right) \cos kx {\rm d}k + \mathcal{O}({\epsilon^2}).
\end{equation}
Eq.~(\ref{convection}) is also written as
\begin{equation}
V(x) = \frac{\epsilon}{2\eta} \int_0^\infty \tilde{c}_0(k) \sin kx {\rm d}k + \mathcal{O}(\epsilon^2).
\end{equation}
At the order of $\epsilon^0$,
\begin{equation}
-D k^2 \tilde{c}_0(k) - a \tilde{c}_0(k) + \frac{f_0}{\pi},
\end{equation}
and the solution is obtained as
\begin{equation}
\tilde{c}_0(k) = \frac{f_0}{\pi(a + D k^2)} = \frac{f_0}{\pi a}\frac{1}{1 + k^2 D / a} = \frac{f_0}{\pi a}\frac{1}{1 + {x_0}^2 k^2}.
\end{equation}
At the order of $\epsilon^1$,
\begin{align}
&\frac{\partial}{\partial x} \left[ \frac{\epsilon}{2\eta} \left( \int_0^\infty \tilde{c}_0(k) \sin kx {\rm d}k\right) \left( \int_0^\infty \tilde{c}_0(k) \cos kx {\rm d}k \right) \right]\nonumber \\ &= -D k^2 \epsilon \tilde{c}_1(k) - a \epsilon \tilde{c}_1(k),
\end{align}
This calculation is performed in the same manner as that in Eq.~(\ref{modecoupling}),
and we obtain
\begin{equation}
\tilde{c}_1(k) = - \frac{{f_0}^2}{2 \pi^2 \eta D a^2} \frac{x_0 k \arctan x_0k + \ln (1 + {x_0}^2k^2)}{ \left(1 + {x_0}^2k^2 \right) \left(4 + {x_0}^2k^2 \right)}. \label{comp1}
\end{equation}

\begin{figure}
\begin{center}
\includegraphics{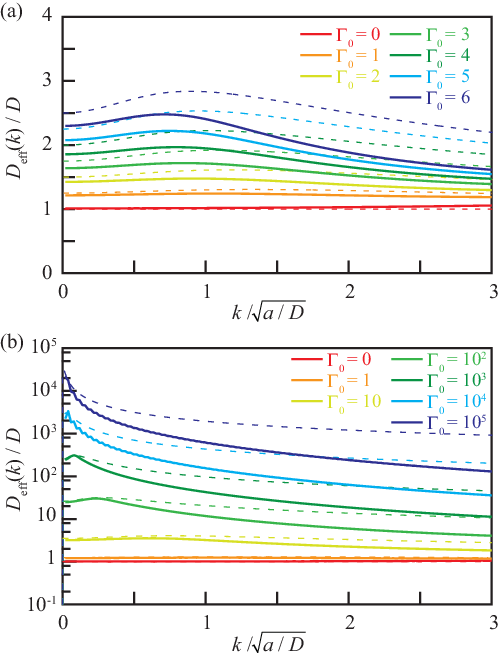}
\end{center}
\caption{Snapshots of concentration and flow field for each time. The parameters are the same as those in Fig.~\ref{fig3}.}
\label{nonlinear}
\end{figure}

\begin{figure}
\begin{center}
\includegraphics{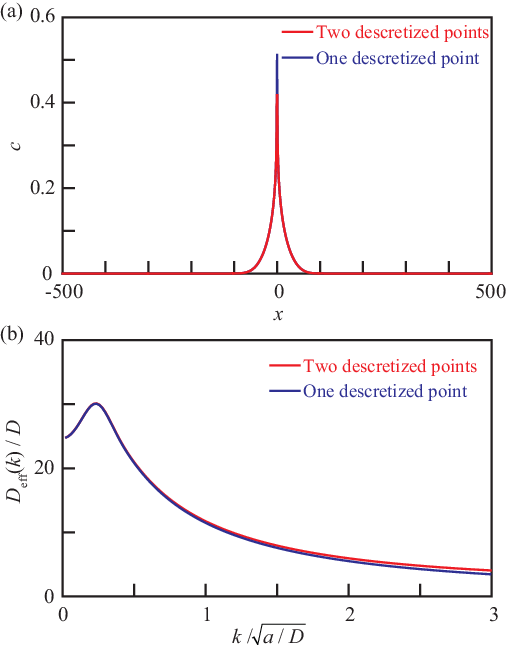}
\end{center}
\caption{Difference between two cases, i.e., the supply at two discretized points (red, the same as the numerical results shown in Figs.~\ref{fig4}(a) and \ref{fig5}(a)) and at one discretized point (blue). (a) Stationary concentration profile of $c$ corresponding to Fig.~\ref{fig4}(a). (b) Plot of $D_{\rm eff}(k) / D$ as a function of $k / \sqrt{a/D}$ corresponding to Fig.~\ref{fig5}(a). $\Gamma_0$ was set to be 100. For the case with the supply at one discretized point, the system size was set to be 1001. The other parameters are the same as those used in Figs.~\ref{fig4} and \ref{fig5}.}
\label{fig_delta}
\end{figure}

\begin{figure}
\begin{center}
\includegraphics{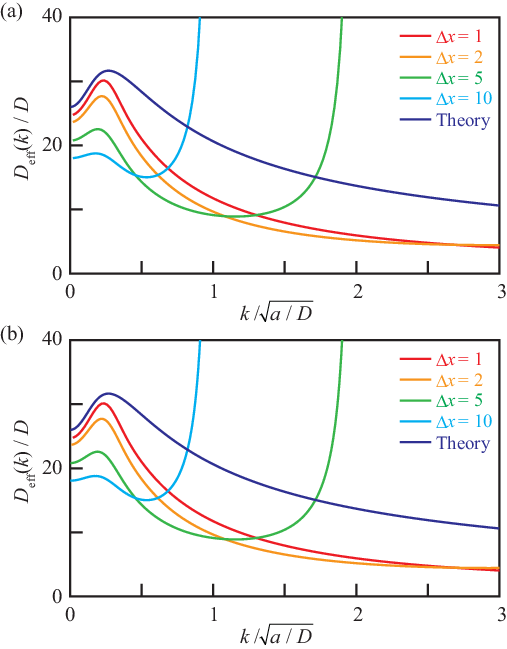}
\end{center}
\caption{Profile of $D_{\rm eff}(k)/D$ against $k / \sqrt{a/D}$ for (a) different $\Delta x$ but with the same system size, and (b) different $\Delta x$ and system size $L_x$ and $L_y$. Here the system size was changed so that $L_x$ and $L_y$ are proportional to $\Delta x$; i.e., $L_x = 500$ and $L_y = 1000$ for $\Delta x = 1$, $L_x = 1000$ and $L_y = 2000$ for $\Delta x = 2$, $L_x = 2500$ and $L_y = 5000$ for $\Delta x = 5$, and $L_x = 5000$ and $L_y = 10000$ for $\Delta x = 10$.  $\Gamma_0$ is set to be 100.}
\label{grid_size}
\end{figure}

\begin{figure}
\begin{center}
\includegraphics{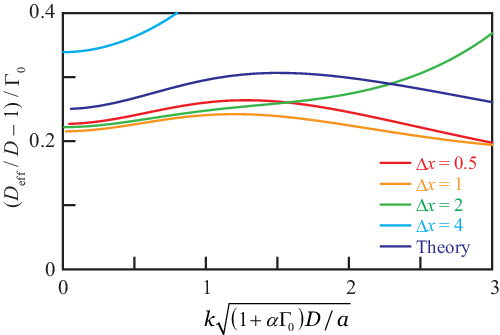}
\end{center}
\caption{Profile of $(D_{\rm eff}(k)/D - 1) / \Gamma_0$ against $k \sqrt{(1 + \alpha \Gamma_0)D/a}$ for different $\Delta x$. The system size was changed so that $L_x$ and $L_y$ are proportional to $\Delta x$; i.e., $L_x = 250$ and $L_y = 500$ for $\Delta x = 0.5$, $L_x = 500$ and $L_y = 1000$ for $\Delta x = 1$, and $L_x = 1000$ and $L_y = 2000$ for $\Delta x = 2$, and $L_x = 2000$ and $L_y = 4000$ for $\Delta x = 4$.  $\Gamma_0$ is set to be 1.}
\label{fig_gridsize_fig6b}
\end{figure}

The perturbed diffusion coefficient $D'$ is also expanded with regard to $\epsilon$ as
\begin{equation}
D' = D'_0 + \epsilon D'_1 + \mathcal{O}(\epsilon^2) = D + \epsilon D'_1 + \mathcal{O}(\epsilon^2).
\end{equation}
Using the relation between $\tilde{c}_n(k)$ and $D'_n$,
\begin{equation}
\tilde{c}_0(k) + \epsilon \tilde{c}_1(k) + \mathcal{O}(\epsilon^2) = \frac{f_0}{\pi} \frac{1}{(D'_0 + \epsilon D'_1)k^2 + a} + \mathcal{O}(\epsilon^2),
\end{equation}
we obtain
\begin{equation}
\tilde{c}_1(k) = - \frac{f_0}{\pi a} \frac{{x_0}^2 k^2}{\left(1 + {x_0}^2k^2 \right)^2} \frac{D'_1}{D} + \mathcal{O}(\epsilon^2). \label{comp2}
\end{equation}
From Eqs.~(\ref{comp1}) and (\ref{comp2}), we derive
\begin{align}
\frac{D'_1}{D} =& \frac{f_0}{2 \pi \eta D a} \frac{1 + {x_0}^2k^2}{{x_0}^2 k^2 (4 + {x_0}^2 k^2 )} \nonumber \\
& \times \left\{ x_0 k \arctan x_0k + \ln (1 + {x_0}^2k^2) \right\} \nonumber \\
=& \frac{f_0}{2 \pi \eta D a} G(x_0 k).
\end{align}
Thus Eq.~(\ref{perturbation}) is obtained.

\section{Confirmation of the validity of our numerical calculation
\label{app-validity}}

\subsection{Effect of nonlinear term in Navier-Stokes equation}
In order to check the importance of the nonlinear term in Navier-Stokes equation, we performed numerical calculation using the Navier-Stokes equation~\eqref{NS} by omitting the nonlinear term $(\bm{v}\cdot \nabla) \bm{v}$. The results corresponding to Fig.~\ref{fig5} are shown in Fig.~\ref{nonlinear}. The profile of concentration field and $D_{\rm eff}$ did not show significant difference from those in Fig.~\ref{fig5}. Therefore we expect the effect of the nonlinear term in the Navier-Stokes equation plays a minor role on the effective diffusion.

\subsection{Effect of the grid number of supplied region}

In order to check the effect of the number of discretized points for the camphor molecule supply, we have calculated the stationary concentration field and the effective diffusion coefficient when the supply of the camphor molecules was at one discretized point. The total amount of the supply was the same, and the system size was set to be 1001. The results are shown in Fig.~\ref{fig_delta}. The concentration field was almost the same except near the supplied region. As for $D_{\rm eff}(k)$, the features did not change at lower $k$ near the peak, but it changed for higher $k$. This indicates the number of discretized number of the supply do not affect the effective diffusion coefficient, though the profile of $D_{\rm eff}$ was changed at high-$k$ region through the concentration profile near the supplied region.

\subsection{Effect of the grid size}

In order to check the source of discrepancy, we performed numerical calculation by changing the spatial mesh $\Delta x$ and also by changing the spatial mesh size $\Delta x$ and system size $L_x$ and $L_y$. The results are shown in Fig.~\ref{grid_size}. In both cases, the profiles of $D_{\rm eff}$ were almost the same, and the profile approached the theoretical prediction as $\Delta x$ became smaller. For greater $\Delta x$, the fine structure around the source was lost and $D_{\rm eff}(k)$ with higher $k$ became greater. Therefore the discrepancy at higher $k$ seems to originate from the discretization of the mesh. In spite of the discrepancy at higher $k$, the peak position near $k = k_{\rm max}$ did not change so significantly. This suggests the validity of the numerical calculation. We have also checked the effect of the grid size for Fig. 6(b). The results are shown in Fig.~\ref{fig_gridsize_fig6b}. For smaller grid size, the curve is closer to the theoretical curve. Therefore, we guess the discrepancies seen in Fig.~\ref{fig6}(b) and (c) are due to the effect of the discretization.


\begin{thebibliography}{99}

\bibitem{Mikhailov}A.~S.~Mikhailov and V.~Calenbuhr, {\it From Cells to Societies} (Springer, Berlin, 2002).

\bibitem{Ramaswarmy}S.~Ramaswamy, Annu. Rev. Cond. Mat. Phys. {\bf 1}, 323 (2010).

\bibitem{Vicsek}T.~Vicsek and A.~Zafeiris, Phys. Rep. {\bf 517}, 71 (2012).

\bibitem{camphor}C.~Tomlinson, Proc. R. Soc. London {\bf 11}, 575 (1860).

\bibitem{Nakata}S.~Nakata, Y.~Iguchi, S.~Ose, M.~Kuboyama, T.~Ishii, and K.~Yoshikawa, Langmuir {\bf 13}, 4454 (1997).

\bibitem{Hayashima}Y.~Hayashima, M.~Nagayama, and S.~Nakata, J. Phys. Chem. B {\bf 105}, 5353 (2001).

\bibitem{PhysicaD}H.~Kitahata and K.~Yoshikawa, Physica D {\bf 205}, 283 (2005).

\bibitem{Grzybowski}S.~Soh , K.~J.~M.~Bishop, and B.~A.~Grzybowski, J. Phys. Chem. B {\bf 112}, 10848 (2008).

\bibitem{Grzybowski2}S.~Soh , M.~Branicki, and B.~A.~Grzybowski, J. Phys. Chem. Lett. {\bf 2}, 770 (2011).

\bibitem{Nishimori1}E.~Heisler, N.~J.~Suematsu, A.~Awazu, and H.~Nishimori, J. Phys. Soc. Jpn. {\bf 81}, 074605 (2012).

\bibitem{Nishimori2}E.~Heisler, N.~J.~Suematsu, A.~Awazu, and H.~Nishimori, Phys. Rev. E {\bf 85}, 055201 (2012).

\bibitem{PCCPreview}S.~Nakata, M.~Nagayama, H.~Kitahata, N.~J.~Suematsu, and T.~Hasegawa, Phys. Chem. Chem. Phys. {\bf 17}, 10326 (2015).

\bibitem{Koyano1}Y.~Koyano, T.~Sakurai, and H.~Kitahata, Phys. Rev. E {\bf 94}, 042215 (2016).

\bibitem{Koyano2}Y.~Koyano, M.~Gryciuk, P.~Skrobanska, M.~Malecki, Y.~Sumino, H.~Kitahata, and J.~Gorecki, Phys. Rev. E {\bf 96}, 012609 (2017).

\bibitem{Nagayama}M.~Nagayama, S.~Nakata, Y.~Doi, and Y.~Hayashima, Physica D {\bf 194}, 151 (2004).

\bibitem{Suematsu}N.~J.~Suematsu, T.~Sasaki, S.~Nakata, and H.~Kitahata, Langmuir {\bf 30}, 8101 (2014).

\bibitem{Marangoni}L.~E.~Scriven and C.~V.~Sternling, Nature {\bf 187}, 186 (1960).

\bibitem{Marangoni2}H.~Linde, P.~Schwartz and H.~Wilke, Dissipative Structures and Nonlinear Kinetics of the Marangoni-Instability, in {\it Dynamics and Instability of Fluid Interfaces}, ed. T.~S.~S{\o}rensen, (Springer-Verlag, Berlin, 1979).

\bibitem{PCCP}H.~Kitahata, S.~Hiromatsu, Y.~Doi, S.~Nakata, and M.~R.~Islam, Phys. Chem. Chem. Phys. {\bf 6}, 2409 (2004).

\bibitem{SDS}Y.~S.~Ikura, R.~Tenno, H.~Kitahata, N.~J.~Suematsu, and S.~Nakata, J. Phys. Chem. B {\bf 116}, 992 (2012). 

\bibitem{Forster}D.~Forster, D.~R.~Nelson, and M.~J.~Stephen, Phys. Rev. A {\bf 16}, 732 (1977).

\bibitem{Leal}L.~G.~Leal, {\it Advanced Transport Phenomena: Fluid Mechanics and Convective Transport Processes} (Cambridge University Press, 2007).

\bibitem{Bratukhin:1968}Y.~K.~Bratukhin and L.~N.~Maurin, J. Eng. Phys. {\bf 14}, 533 (1968).

\bibitem{Roche:2014}M.~Roch{\'{e}}, Z.~Li, I.~M.~Griffiths, S.~Le Roux, I.~Cantat, A.~Saint-Jalmes, and H.~A.~Stone, Phys. Rev. Lett. {\bf 112}, 208302 (2014).

\bibitem{LeRoux:2016}S.~Le Roux, M.~Roch{\'{e}}, I.~Cantat, and A.~Saint-Jalmes, Phys. Rev. E {\bf 93}, 013107 (2016).

\bibitem{Mandre:2017b}M.~M.~Bandi, V.~S.~Akella, D.~K.~Singh, R.~S.~Singh, and S.~Mandre, Phys. Rev. Lett. {\bf 119}, 264501 (2017).

\bibitem{Mandre:2017}S.~Mandre, J. Fluid Mech. {\bf 832} 777 (2017).

\bibitem{interface}A.~A.~Nepomniashchy, M.~G.~Velarde, P.~Colinet, {\it Interfacial Phenomena and Convection} (Chapman \& Hall/CRC, Boca Raton, 2002).

\bibitem{Landau}L.~D.~Landau and E.~M.~Lifshitz, {\it Fluid Mechanics} (Pergamon Press, Oxford, 1959).

\bibitem{Young}N.~O.~Young, J.~S.~Goldstein, and M.~J.~Block, J. Fluid Mech. {\bf 6}, 350 (1959).

\bibitem{Brenner}J.~Happel and H.~Brenner, {\it Low Reynolds Number Hydrodynamics: With Special Applications to Particulate Media} (Prentice-Hall, Englewood Cliffs, 1965).

\bibitem{Chang}C.-H. Chang, E.~I.~Franses, Colloids Surfaces A {\bf 100}, 1 (1995).

\bibitem{Kitahata:2017}H.~Kitahata, H.~Yamamoto, M.~Hata, Y.~S.~Ikura, and S.~Nakata, Colloids Surfaces A, {\bf 520}, 436 (2017).

\end{thebibliography}
\end{document}